\documentclass[aps,prl,preprint,tightenlines,superscriptaddress,showpacs,byrevtex,floatfix]{revtex4}
\usepackage{graphicx}
\usepackage{dcolumn}
\usepackage{bm}

\newcommand*{\sinbb}{{\sin2\phi_1}}
\newcommand*{\dM}{\ensuremath{{\Delta m_d}}}

\newcommand*{\bz}{\ensuremath{{B^0}}}
\newcommand*{\bzb}{\ensuremath{{\overline{B}{}^0}}}
\newcommand*{\bp}{\ensuremath{{B^+}}}

\newcommand*{\piz}{\ensuremath{{\pi^0}}}
\newcommand*{\pip}{\ensuremath{{\pi^+}}}
\newcommand*{\pim}{\ensuremath{{\pi^-}}}
\newcommand*{\kp}{\ensuremath{{K^+}}}
\newcommand*{\km}{\ensuremath{{K^-}}}
\newcommand*{\ks}{\ensuremath{{K_S^0}}}
\newcommand*{\kl}{\ensuremath{{K_L^0}}}
\newcommand*{\kstarz}{\ensuremath{{K^{*0}}}}
\newcommand*{\jpsi}{\ensuremath{{J/\psi}}}

\newcommand*{\Dt}{\ensuremath{{\Delta t}}}
\newcommand*{\Dz}{\ensuremath{{\Delta z}}}

\newcommand*{\fcp}{\ensuremath{{f_{CP}}}}
\newcommand*{\ftag}{\ensuremath{{f_\textrm{tag}}}}
\newcommand*{\tcp}{\ensuremath{{t_{CP}}}}
\newcommand*{\ttag}{\ensuremath{{t_\textrm{tag}}}}
\newcommand*{\zcp}{\ensuremath{{z_{CP}}}}
\newcommand*{\ztag}{\ensuremath{{z_\textrm{tag}}}}

\newcommand*{\dE}{\ensuremath{{\Delta E}}}
\newcommand*{\mb}{\ensuremath{{M_\textrm{bc}}}}
\newcommand*{\Ebeam}{\ensuremath{{E_\textrm{beam}^\textrm{cms}}}}
\newcommand*{\EB}{\ensuremath{{E_B^\textrm{cms}}}}
\newcommand*{\pB}{\ensuremath{{p_B^\textrm{cms}}}}

\newcommand*{\Nev}{\ensuremath{{N_\textrm{ev}}}}

\newcommand*{\eeff}{\ensuremath{\epsilon_\textrm{eff}}}

\newcommand*{\taubz}{\ensuremath{{\tau_\bz}}}
\newcommand*{\taubp}{\ensuremath{{\tau_\bp}}}

\newcommand*{\Psig}{\ensuremath{{\mathcal{P}_\textrm{sig}}}}
\newcommand*{\Pbkg}{\ensuremath{{\mathcal{P}_\textrm{bkg}}}}
\newcommand*{\Pol}{\ensuremath{{P_\textrm{ol}}}}
\newcommand*{\Rsig}{\ensuremath{{R_\textrm{sig}}}}
\newcommand*{\Rbkg}{\ensuremath{{R_\textrm{bkg}}}}
\newcommand*{\fol}{\ensuremath{{f_\textrm{ol}}}}
\newcommand*{\fsig}{\ensuremath{{f_\textrm{sig}}}}

\newcommand*{\dwl}{\ensuremath{{\Delta w_l}}}
\newcommand*{\fq}{\ensuremath{q}}

\newcommand{\cala}{{\cal A}}
\newcommand{\cals}{{\cal S}}

\newcommand*{\sinbbcenter}{0.733}
\newcommand*{\sinbbstat}{\pm0.057}
\newcommand*{\sinbbsys}{\pm0.028}
\newcommand*{\sinbbresult}{\sinbbcenter\sinbbstat \textrm{(stat)} \sinbbsys \textrm{(syst)}}
\newcommand*{\efftot}{0.287 \pm 0.005}
\newcommand*{\lambdacenter}{1.007}
\newcommand*{\lambdastat}{\pm0.041}
\newcommand*{\lambdaresult}{\lambdacenter\lambdastat\textrm{(stat)}}

\newcommand*{\nevqp}{2717}
\newcommand*{\nevqm}{2700}

\begin{document}

\vspace*{-3\baselineskip}
\resizebox{!}{3cm}{\includegraphics{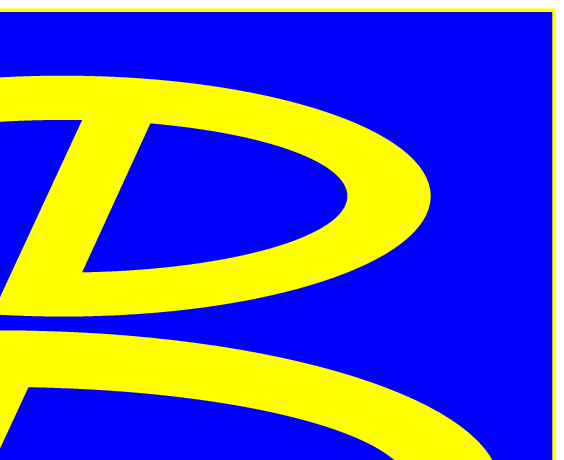}}

\preprint{BELLE-CONF-0353}

\title{Measurement of
{\boldmath $CP$}-Violation Parameter {\boldmath $\sin2\phi_1$}\\ 
with 152 Million {\boldmath $B\overline{B}$} Pairs}

\date{\today}

\affiliation{Aomori University, Aomori}
\affiliation{Budker Institute of Nuclear Physics, Novosibirsk}
\affiliation{Chiba University, Chiba}
\affiliation{Chuo University, Tokyo}
\affiliation{University of Cincinnati, Cincinnati, Ohio 45221}
\affiliation{University of Frankfurt, Frankfurt}
\affiliation{Gyeongsang National University, Chinju}
\affiliation{University of Hawaii, Honolulu, Hawaii 96822}
\affiliation{High Energy Accelerator Research Organization (KEK), Tsukuba}
\affiliation{Hiroshima Institute of Technology, Hiroshima}
\affiliation{Institute of High Energy Physics, Chinese Academy of Sciences, Beijing}
\affiliation{Institute of High Energy Physics, Vienna}
\affiliation{Institute for Theoretical and Experimental Physics, Moscow}
\affiliation{J. Stefan Institute, Ljubljana}
\affiliation{Kanagawa University, Yokohama}
\affiliation{Korea University, Seoul}
\affiliation{Kyoto University, Kyoto}
\affiliation{Kyungpook National University, Taegu}
\affiliation{Institut de Physique des Hautes \'Energies, Universit\'e de Lausanne, Lausanne}
\affiliation{University of Ljubljana, Ljubljana}
\affiliation{University of Maribor, Maribor}
\affiliation{University of Melbourne, Victoria}
\affiliation{Nagoya University, Nagoya}
\affiliation{Nara Women's University, Nara}
\affiliation{National Kaohsiung Normal University, Kaohsiung}
\affiliation{National Lien-Ho Institute of Technology, Miao Li}
\affiliation{Department of Physics, National Taiwan University, Taipei}
\affiliation{H. Niewodniczanski Institute of Nuclear Physics, Krakow}
\affiliation{Nihon Dental College, Niigata}
\affiliation{Niigata University, Niigata}
\affiliation{Osaka City University, Osaka}
\affiliation{Osaka University, Osaka}
\affiliation{Panjab University, Chandigarh}
\affiliation{Peking University, Beijing}
\affiliation{Princeton University, Princeton, New Jersey 08545}
\affiliation{RIKEN BNL Research Center, Upton, New York 11973}
\affiliation{Saga University, Saga}
\affiliation{University of Science and Technology of China, Hefei}
\affiliation{Seoul National University, Seoul}
\affiliation{Sungkyunkwan University, Suwon}
\affiliation{University of Sydney, Sydney NSW}
\affiliation{Tata Institute of Fundamental Research, Bombay}
\affiliation{Toho University, Funabashi}
\affiliation{Tohoku Gakuin University, Tagajo}
\affiliation{Tohoku University, Sendai}
\affiliation{Department of Physics, University of Tokyo, Tokyo}
\affiliation{Tokyo Institute of Technology, Tokyo}
\affiliation{Tokyo Metropolitan University, Tokyo}
\affiliation{Tokyo University of Agriculture and Technology, Tokyo}
\affiliation{Toyama National College of Maritime Technology, Toyama}
\affiliation{University of Tsukuba, Tsukuba}
\affiliation{Utkal University, Bhubaneswer}
\affiliation{Virginia Polytechnic Institute and State University, Blacksburg, Virginia 24061}
\affiliation{Yokkaichi University, Yokkaichi}
\affiliation{Yonsei University, Seoul}
  \author{K.~Abe}\affiliation{High Energy Accelerator Research Organization (KEK), Tsukuba} 
  \author{K.~Abe}\affiliation{Tohoku Gakuin University, Tagajo} 
  \author{N.~Abe}\affiliation{Tokyo Institute of Technology, Tokyo} 
  \author{R.~Abe}\affiliation{Niigata University, Niigata} 
  \author{T.~Abe}\affiliation{High Energy Accelerator Research Organization (KEK), Tsukuba} 
  \author{I.~Adachi}\affiliation{High Energy Accelerator Research Organization (KEK), Tsukuba} 
  \author{Byoung~Sup~Ahn}\affiliation{Korea University, Seoul} 
  \author{H.~Aihara}\affiliation{Department of Physics, University of Tokyo, Tokyo} 
  \author{M.~Akatsu}\affiliation{Nagoya University, Nagoya} 
  \author{M.~Asai}\affiliation{Hiroshima Institute of Technology, Hiroshima} 
% ...
\collaboration{The Belle Collaboration}

\begin{abstract}
We present a precise measurement of the standard model $CP$-violation
parameter $\sinbb$ based on a sample
of $152 \times 10^6$ $B\overline{B}$ pairs
collected at the $\Upsilon(4S)$ resonance
with the Belle detector at the KEKB asymmetric-energy $e^+e^-$ collider.
One neutral $B$ meson is reconstructed in a $\jpsi\ks$, $\psi(2S)\ks$,
$\chi_{c1}\ks$, $\eta_c\ks$, $\jpsi\kstarz$, or $\jpsi\kl$ $CP$-eigenstate
decay channel and the flavor of the accompanying $B$ meson is
identified from its decay products.
From the asymmetry in the distribution of the time interval
between the two $B$ meson decay points, we obtain
$\sinbb = \sinbbresult$.
\end{abstract}


\maketitle

In the standard model (SM), $CP$ violation arises from an
irreducible phase in the weak interaction quark-mixing matrix
[Cabibbo-Kobayashi-Maskawa (CKM) matrix]~\cite{bib:ckm}.
In particular, the SM predicts a $CP$-violating asymmetry
in the time-dependent rates for $\bz$ and $\bzb$ decays
to a common $CP$ eigenstate $\fcp$,
where the transition is dominated by the $b \to c\overline{c}s$ process,
with negligible corrections from strong interactions~\cite{bib:sanda}:
\begin{equation}
  A(t) \equiv \frac{\Gamma(\bzb \to \fcp) - \Gamma(\bz \to \fcp)}
  {\Gamma(\bzb \to \fcp) + \Gamma(\bz \to \fcp)}
  = -\xi_f \sinbb \sin(\dM t),
\end{equation}
where $\Gamma(\bz,\bzb \to \fcp)$ is the rate for $\bz$ or $\bzb$
to $\fcp$ at a proper time $t$ after production,
$\xi_f$ is the $CP$ eigenvalue of $\fcp$,
$\dM$ is the mass difference between the two $\bz$ mass eigenstates,
and $\phi_1$ is one of the three interior angles of the CKM unitarity triangle,
defined as $\phi_1 \equiv \pi - \arg(V_{tb}^*V_{td}/V_{cb}^*V_{cd})$.
Non-zero values for $\sinbb$ have been reported
by the Belle and BaBar collaborations~\cite{bib:cpv,bib:Belle_sin2phi1_78fb-1,bib:babar}.

Belle's latest published measurement of $\sinbb$ is based on
a 78~fb$^{-1}$ data sample (data set I) containing $85 \times 10^{6}$
$B\overline{B}$ pairs produced at the $\Upsilon(4S)$ resonance.
In this paper, we report an improved measurement 
incorporating an additional 62 fb$^{-1}$ (data set II) for a total of
140 fb$^{-1}$ ($152 \times 10^6$ $B\overline{B}$ pairs).
Changes exist in the analysis
with respect to our earlier result~\cite{bib:Belle_sin2phi1_78fb-1}.
We apply a new proper-time interval resolution function
that reduces systematic uncertainties in $\sinbb$ and also in 
$\dM$ and lifetime ($\taubz$, $\taubp$) measurements.
We introduce $b$-flavor-dependent wrong-tag fractions
to accommodate possible differences between $\bz$ and $\bzb$ decays.
We also adopt a multi-parameter fit to the flavor-specific control samples
to obtain the resolution parameters and wrong-tag fractions simultaneously.
There are other improvements in the estimation of background
components, which become possible with increased statistics.

The data were collected with the Belle detector~\cite{bib:belle}
at the KEKB  asymmetric collider~\cite{bib:KEKB},
which collides 8.0~GeV $e^-$ on 3.5~GeV $e^+$
at a small ($\pm 11$~mrad) crossing angle.
We use events where one of the $B$ mesons decays to $\fcp$ at time $\tcp$,
and the other decays to a self-tagging state $\ftag$,
which distinguishes $\bz$ from $\bzb$, at time $\ttag$.
The $CP$ violation manifests itself as an asymmetry $A(\Dt)$,
where $\Dt$ is the proper time interval
between the two decays: $\Dt \equiv \tcp - \ttag$.
At KEKB, the $\Upsilon(4S)$ resonance is produced
with a boost of $\beta\gamma = 0.425$ nearly along the $z$ axis
defined as anti-parallel to the positron beam direction,
and $\Dt$ can be determined as $\Dt \simeq \Dz/(\beta\gamma)c$,
where $\Dz$ is the $z$ distance between the $\fcp$ and $\ftag$
decay vertices, $\Dz \equiv \zcp - \ztag$.
The average value of $\Dz$ is approximately 200~$\mu$m.

The Belle detector~\cite{bib:belle} is a large-solid-angle spectrometer
that includes a silicon vertex detector (SVD),
a central drift chamber (CDC),
an array of aerogel threshold \v{C}erenkov counters (ACC),
time-of-flight (TOF) scintillation counters,
and an electromagnetic calorimeter comprised of CsI(Tl) crystals (ECL)
located inside a superconducting solenoid coil
that provides a 1.5~T magnetic field.
An iron flux-return located outside of the coil is instrumented
to detect $\kl$ mesons and to identify muons (KLM).

We reconstruct $\bz$ decays to the following $CP$ 
eigenstates~\cite{footnote:cc}:
$\jpsi\ks$, $\psi(2S)\ks$, $\chi_{c1}\ks$, $\eta_c\ks$ for $\xi_f = -1$
and $\jpsi\kl$ for $\xi_f = +1$.
We also use $\bz \to \jpsi\kstarz$ decays where $\kstarz \to \ks\piz$.
Here the final state is a mixture of even and odd $CP$,
depending on the relative orbital angular momentum of the $\jpsi$ and $\kstarz$.
We find that the final state is primarily $\xi_f = +1$;
the $\xi_f = -1$ fraction is
$0.19 \pm 0.02 \textrm{(stat)} \pm 0.03 \textrm{(syst)}$~\cite{bib:itoh}.

The reconstruction and selection criteria for all $\fcp$ channels
used in the measurement are described in detail elsewhere~\cite{bib:cpv}.
$\jpsi$ and $\psi(2S)$ mesons are reconstructed
via their decays to $\ell^+\ell^-$ ($\ell = \mu,e$).
The $\psi(2S)$ is also reconstructed via $\jpsi\pip\pim$,
and the $\chi_{c1}$ via $\jpsi\gamma$.
The $\eta_c$ is detected in the $\ks\km\pip$, $\kp\km\piz$,
and $p\overline{p}$ modes.
For the $\jpsi\ks$ mode, we use $\ks \to \pip\pim$ and $\piz\piz$ decays;
for other modes we only use $\ks \to \pip\pim$.
For reconstructed $B \to \fcp$ candidates other than $\jpsi\kl$,
we identify $B$ decays using the energy difference $\dE \equiv \EB - \Ebeam$
and the beam-energy constrained mass $\mb \equiv \sqrt{(\Ebeam)^2-(\pB)^2}$,
where $\Ebeam$ is the beam energy in the center-of-mass system (cms)
of the $\Upsilon(4S)$ resonance, and $\EB$ and $\pB$ are
the cms energy and momentum of the reconstructed $B$ candidate, respectively.

Candidate $\bz \to \jpsi\kl$ decays are selected by requiring
ECL and/or KLM hit patterns that are consistent with the presence
of a shower induced by a $\kl$ meson.
The centroid of the shower is required to be within a $45^\circ$ cone
centered on the $\kl$ direction inferred from
two-body decay kinematics and the measured four-momentum of the $\jpsi$.

We perform a multi-parameter fit
to flavor-specific control samples
to obtain wrong-tag fractions and parameters for the resolution function simultaneously.
We select $\bz \to D^{*-}\ell^+\nu$, $\jpsi\kstarz(\kp\pim)$,
$D^{*-}\pi^+$, $D^-\pi^+$, $D^{*-}\rho^+$, and
$J/\psi \ks(\ell^+\ell^-)$ (for resolution parameters only)
for $\bz$ decays,
and $B^+ \to \overline{D}{}^0\pi^+$ and $J/\psi K^+$
for $B^+$ decays.
The total numbers of candidates ($\Nev$) and purities ($p$) are
$\Nev = 124118$ and $p = 0.82$ for $B^0$ decays, and
$\Nev = 57305$ and $p = 0.81$ for $B^+$ decays.
The fit uses free parameters for wrong-tag fractions (12), 
for the resolution function (14),
for the $B^+$ background in $\bz$ decays (3),
$\dM$, $\taubz$ and $\taubp$.
The total number of parameters in the fit is 32.
We add two parameters to the resolution function
described in~\cite{bib:resol}
to obtain an improved description of the
effect of charmed particle decays in the $\ftag$ vertex.
We test the new fit method and parameterization with 
a large number of Monte Carlo (MC) events.
A fit to the MC control sample
yields $\dM = (0.488\pm0.002)$ ps$^{-1}$, 
$\taubz = (1.539\pm 0.003)$ ps and
$\taubp = (1.679\pm 0.004)$ ps for the input values of
$\dM = 0.489$ ps$^{-1}$,
$\taubz = 1.541$ ps and
$\taubp = 1.674$ ps, respectively.
The obtained wrong-tag fractions are also found to be correct. 
The unbinned maximum-likelihood fit to data yields 
$\dM = [0.511\pm0.005$(stat)] ps$^{-1}$, 
$\taubz = [1.533\pm0.008$(stat)] ps and 
$\taubp = [1.634\pm0.011$(stat)] ps, 
where the errors are statistical only.
The results are consistent with the present world average 
values~\cite{bib:PDG2003}.

Charged leptons, pions, kaons, and $\Lambda$ baryons
that are not associated with a reconstructed $CP$ eigenstate decay
are used to identify the $b$-flavor of the accompanying $B$ meson.
The tagging algorithm is identical to the one used
in reference~\cite{bib:Belle_sin2phi1_78fb-1}.
We use two parameters, $\fq$ and $r$, to represent the tagging information.
The first, $q$, has the discrete value $+1$~($-1$)
when the tag-side $B$ meson is likely to be a $\bz$~($\bzb$),
and the parameter $r$ is an event-by-event Monte Carlo-determined
flavor-tagging dilution parameter that ranges
from $r=0$ for no flavor discrimination
to $r=1$ for an unambiguous flavor assignment.
It is used only to sort data into six intervals of $r$,
according to estimated flavor purity.
We determine directly from data
the average wrong-tag probabilities, 
$w_l \equiv (w_l^+ + w_l^-)/2~(l=1,6)$,
and differences between $\bz$ and $\bzb$ decays, 
$\dwl \equiv w_l^+ - w_l^-$,
where $w_l^{+(-)}$ is the wrong-tag probability
for the $\bz(\bzb)$ decay in each $r$ interval. 
The event fractions and wrong-tag fractions
are summarized in Table~\ref{tab:wtag}.
\begin{table}
  \caption{\label{tab:wtag} The event fractions $\epsilon_l$,
    wrong-tag fractions $w_l$, wrong-tag fraction differences $\dwl$,
    and average effective tagging efficiencies
    $\eeff^l = \epsilon_l(1-2w_l)^2$ for each $r$ interval.
    The errors include both statistical and systematic uncertainties.
    The event fractions are obtained from the $\jpsi\ks$ simulation.}
  \begin{ruledtabular}
    \begin{tabular}{ccclll}
      $l$ & $r$ interval & $\epsilon_l$ &\multicolumn{1}{c}{$w_l$} 
          & \multicolumn{1}{c}{$\dwl$}  &\multicolumn{1}{c}{$\eeff^l$} \\
      \hline
 1 & 0.000 -- 0.250 & 0.398 & $0.464\pm0.006$ &$-0.011\pm0.006$ &$0.002\pm0.001$ \\
 2 & 0.250 -- 0.500 & 0.146 & $0.331\pm0.008$ &$+0.004\pm0.010$ &$0.017\pm0.002$ \\
 3 & 0.500 -- 0.625 & 0.104 & $0.231\pm0.009$ &$-0.011\pm0.010$ &$0.030\pm0.002$ \\
 4 & 0.625 -- 0.750 & 0.122 & $0.163\pm0.008$ &$-0.007\pm0.009$ &$0.055\pm0.003$ \\
 5 & 0.750 -- 0.875 & 0.094 & $0.109\pm0.007$ &$+0.016\pm0.009$ &$0.057\pm0.002$ \\
 6 & 0.875 -- 1.000 & 0.136 & $0.020\pm0.005$ &$+0.003\pm0.006$ &$0.126\pm0.003$ \\
    \end{tabular}
  \end{ruledtabular}
\end{table}
The total effective tagging efficiency is determined to be
$\eeff \equiv \sum_{l=1}^6 \epsilon_l(1-2w_l)^2 = \efftot$,
where $\epsilon_l$ is the event fraction for each $r$ interval.
The error includes both statistical and systematic uncertainties.

The vertex position for the $\fcp$ decay is reconstructed
using leptons from $\jpsi$ decays or charged hadrons from $\eta_c$ decays,
and that for $\ftag$ is obtained with well reconstructed tracks
that are not assigned to $\fcp$.
Tracks that are consistent with coming from a $\ks\to\pip\pim$ decay
are not used.
Each vertex position is required to be consistent with
the interaction region profile, determined run-by-run,
smeared in the $r$-$\phi$ plane to account for the $B$ meson decay length.
With these requirements, we are able to determine a vertex
even with a single track;
the fraction of single-track vertices is about 10\% for $\zcp$
and 22\% for $\ztag$.
The proper-time interval resolution function $\Rsig(\Dt)$
is formed by convolving four components:
the detector resolutions for $\zcp$ and $\ztag$,
the shift in the $\ztag$ vertex position
due to secondary tracks originating from charmed particle decays,
and the kinematic approximation that the $B$ mesons are
at rest in the cms~\cite{bib:resol}.
A small component of broad outliers in the $\Dz$ distribution,
caused by mis-reconstruction, is represented by a Gaussian function.
We determine fourteen resolution parameters from the
aforementioned fit to the control samples.
We find that the average $\Dt$ resolution is $\sim 1.43$~ps (rms).
The width of the outlier component
is determined to be $(39\pm 2)$~ps;
the fractions of the outlier components are $(2.1 \pm 0.6) \times 10^{-4}$
for events with both vertices reconstructed with more than one track,
and $(3.1 \pm 0.1) \times 10^{-2}$ for events with at least
one single-track vertex.

\begin{figure}
  \includegraphics[width=0.6\textwidth,clip]{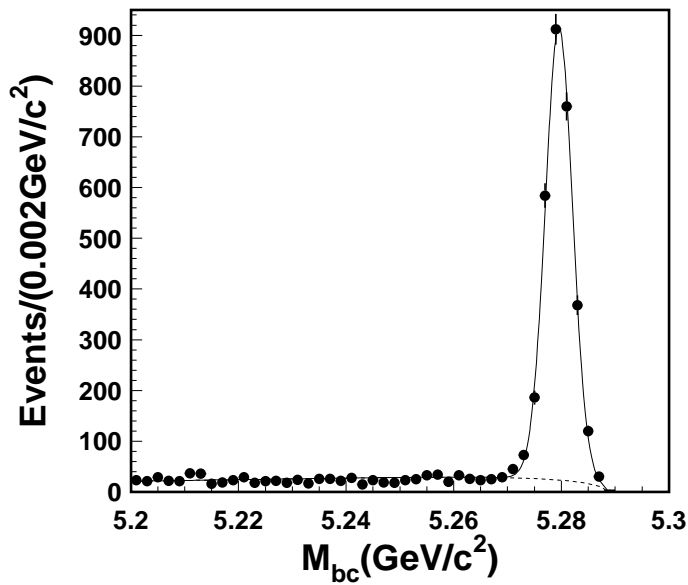}
  \caption{\label{fig:mbc} The beam-energy constrained mass distribution
    within the $\dE$ signal region for all $f_{CP}$ modes other than $\jpsi\kl$. 
    The solid curve shows the fit to signal plus background distributions, and
    the dashed curve shows the background contribution.}
\end{figure}

\begin{figure}
  \includegraphics[width=0.6\textwidth,clip]{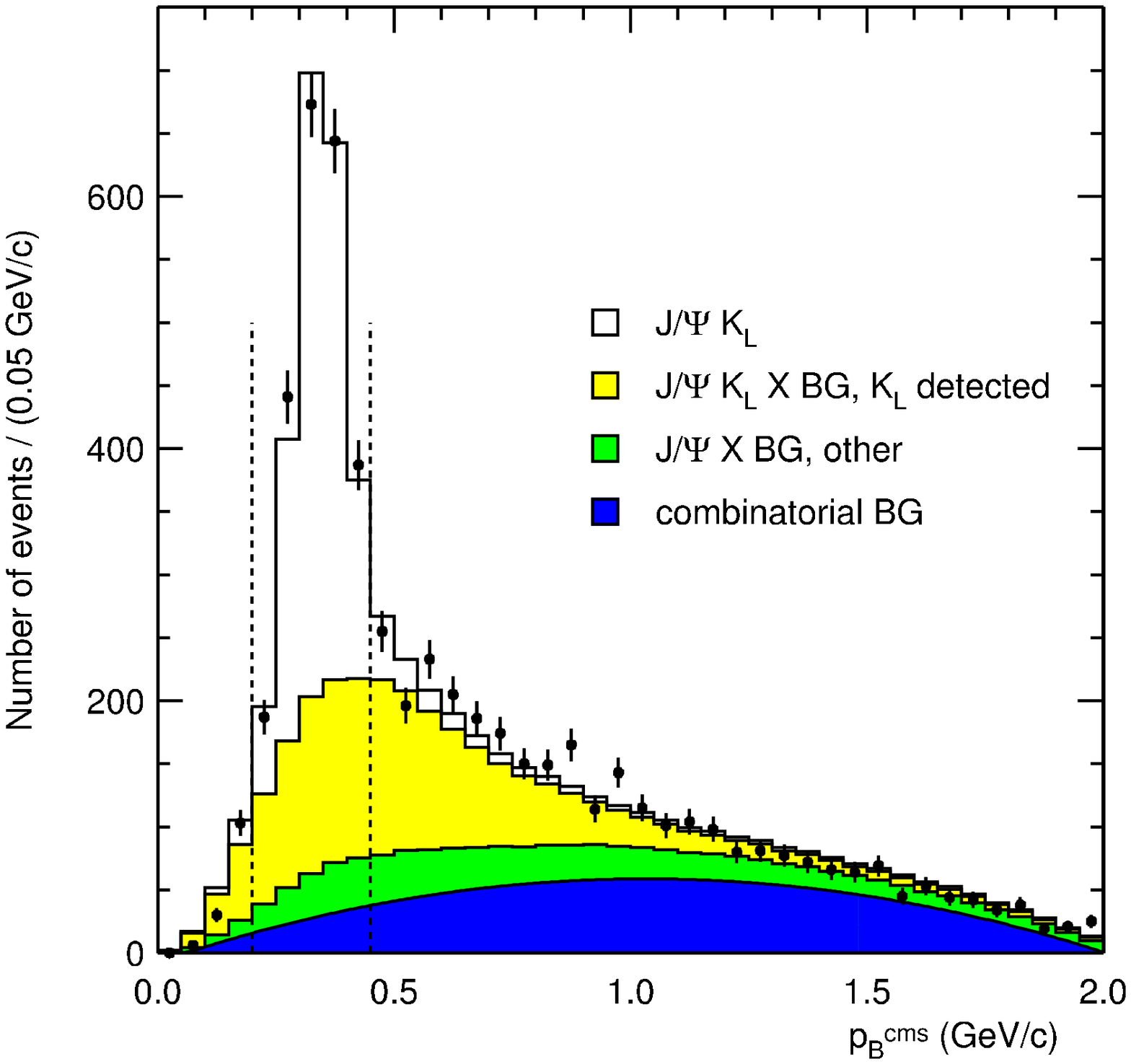}
  \caption{\label{fig:pbstar} 
    The $p_B^\textrm{cms}$ distribution for $\bz \to \jpsi\kl$ candidates
    with the results of the fit. The dashed lines indicate the
    signal region ($0.20 \le \pB \le 0.45$ GeV/$c$).}
\end{figure}

After flavor tagging and vertexing, we find 5417 events in total in the
signal region, which are used for the $\sinbb$ determination.
Table~\ref{tab:number} lists the numbers of candidates, $\Nev$,
and the estimated signal purity for each $\fcp$ mode.
\begin{table}
  \caption{\label{tab:number} The numbers of reconstructed $B \to \fcp$
    candidates after flavor tagging and vertex reconstruction, $\Nev$,
    and the estimated signal purity, $p$, in the signal region for each $\fcp$ mode.
    $\jpsi$ mesons are reconstructed in $\jpsi \to \mu^+\mu^-$ or $e^+e^-$
    decays. Candidate $\ks$ mesons are reconstructed in $\ks \to \pi^+\pi^-$
    decays unless otherwise written explicitly.}
  \begin{ruledtabular}
    \begin{tabular}{llrl}
      \multicolumn{1}{c}{Mode} & $\xi_f$ & $\Nev$ & \multicolumn{1}{c}{$p$} \\
      \hline 
      $J/\psi \ks $                & $-1$ & 1997 & $0.976\pm 0.001$ \\
      $J/\psi \ks(\piz\piz)$       & $-1$ &  288 & $0.82~\pm 0.02$ \\
      $\psi(2S)(\ell^+\ell^-)\ks$  & $-1$ &  145 & $0.93~\pm 0.01$ \\
      $\psi(2S)(\jpsi\pip\pim)\ks$ & $-1$ &  163 & $0.88~\pm 0.01$ \\
      $\chi_{c1}(\jpsi\gamma)\ks$  & $-1$ &  101 & $0.92~\pm 0.01$ \\
      $\eta_c(\ks\km\pip)\ks$      & $-1$ &  123 & $0.72~\pm 0.03$ \\
      $\eta_c(\kp\km\piz)\ks$      & $-1$ &   74 & $0.70~\pm 0.04$ \\
      $\eta_c(p\overline{p})\ks$   & $-1$ &   20 & $0.91~\pm 0.02$ \\
      \cline{3-4}
      All with $\xi_f = -1$        & $-1$ & 2911 & $0.933\pm 0.002$ \\
      \hline
      $J/\psi\kstarz(\ks\piz)$ & +1(81\%)
                                          &  174 & $0.93~\pm 0.01$ \\
      \hline
      $J/\psi\kl$                  & $+1$ & 2332 & $0.60~\pm 0.03$ \\
    \end{tabular}
  \end{ruledtabular}
\end{table}
%
Figure~\ref{fig:mbc} shows the $\mb$ distribution
after applying mode-dependent requirements on $\dE$
for all $\bz$ candidates except for $\bz \to \jpsi\kl$.
There are 3085 entries in total in the signal region defined
as $5.27 < \mb < 5.29$ GeV/$c^2$.
Figure~\ref{fig:pbstar} shows the $\pB$ distribution
for $\bz \to \jpsi\kl$ candidates. We find 2332 entries
in the $0.20 \le \pB \le 0.45$ GeV/$c$ signal region.

\begin{figure}
  \includegraphics[width=0.6\textwidth,clip]{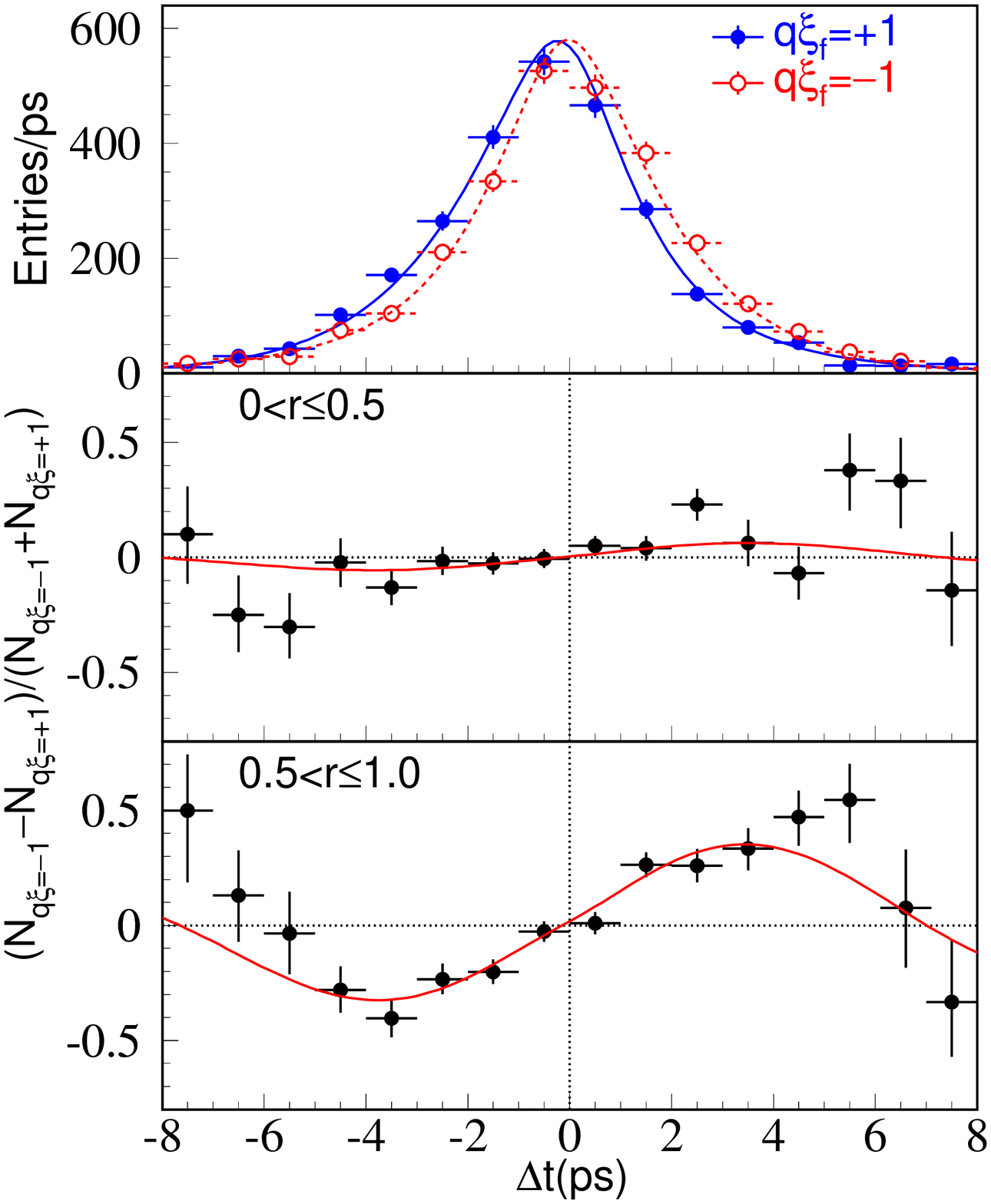}
  \caption{\label{fig:cpfit} The $\Dt$ distributions for the events
    with $q\xi_f = -1$ (open points) and 
    $q\xi_f = +1$ (solid points) with all modes combined (top),
    the asymmetry in each $\Dt$ bin between $q\xi_f=-1$ and $q\xi_f=+1$
    samples with $0 < r \le 0.5$ (middle), and
    with $0.5 < r \le 1$ (bottom).
    The results of the global unbinned maximum-likelihood fit 
    ($\sinbb=0.733$) are also shown.}
\end{figure}

Figure~\ref{fig:cpfit} shows the observed $\Dt$ distributions
for the $q\xi_f = +1$ and $q\xi_f = -1$ event samples (top), 
the asymmetry between two samples with $0 < r \le 0.5$ (middle)
and with $0.5 < r \le 1.0$ (bottom).
The asymmetry in the region $0.5 < r \le 1.0$,
where wrong-tag fractions are small
as shown in Table~\ref{tab:wtag},
clearly demonstrates large $CP$ violation.

We determine $\sinbb$ from an unbinned maximum-likelihood fit
to the observed $\Dt$ distributions.
The probability density function (PDF) expected
for the signal distribution is given by
\begin{equation}
  \label{eq:deltat}
  \Psig(\Dt, q, w_l, \dwl, \xi_f) =
  \frac{e^{-|\Dt|/\taubz}}{4\taubz}[1 - q\dwl -q\xi_f(1-2w_l)\sinbb\sin(\dM\Dt)],
\end{equation}
where we fix the $\bz$ lifetime $\taubz$ and mass difference $\dM$
at their world average values~\cite{bib:PDG2003}.
Each PDF is convolved with the appropriate $\Rsig(\Dt)$
to determine the likelihood value for each event as a function of $\sinbb$:
\begin{eqnarray}
  P_i &=& (1-\fol) \int \Bigl[ \fsig \Psig(\Dt',q,w_l,\dwl,\xi_f)\Rsig(\Dt-\Dt')
  \nonumber \\
  && \quad + \; (1-\fsig)\Pbkg(\Dt')\Rbkg(\Dt-\Dt')\Bigr] d\Dt'
  + \fol \Pol(\Dt),
\end{eqnarray}
where $\fsig$ is the signal fraction
calculated as a function of $\pB$ for $\jpsi\kl$
and of $\dE$ and $\mb$ for other modes.
$\Pbkg(\Dt)$ is the PDF for combinatorial background events,
which is modeled as a sum of exponential and prompt components.
It is convolved with a sum of two Gaussians, $\Rbkg$,
which is regarded as a resolution function for the background.
To account for a small number of events that give large $\Dt$ in
both the signal and background, we introduce
the PDF of the outlier component, $\Pol$, and its fraction $\fol$.
The only free parameter in the final fit is $\sinbb$,
which is determined by maximizing the likelihood function $L = \prod_i P_i$,
where the product is over all events.
The result of the fit is
\[
\sinbb = \sinbbresult .
\]

Sources of systematic error include uncertainties
in the flavor tagging (0.014),
in the vertex reconstruction (0.013),
in the background fractions for
$\bz \to \jpsi\kl$ (0.012) and for other modes (0.007),
in the resolution function (0.008),
a possible bias in the $\sinbb$ fit (0.008), and
an effect of interferences~\cite{bib:fbtginterference} 
in the $f_{\rm tag}$ final state (0.008).
The errors introduced by uncertainties 
in $\dM$, $\taubz$ and in the background $\Dt$ distribution, 
are less than 0.005.

Several checks on the measurement are performed.
Table~\ref{tab:check} lists the results obtained by applying the same analysis
to various subsamples.
\begin{table}
  \caption{\label{tab:check} The numbers of candidate events, $\Nev$,
    and values of $\sinbb$ for various subsamples (statistical errors only).}
  \begin{ruledtabular}
    \begin{tabular}{lrc}
      Sample & \multicolumn{1}{c}{$\Nev$} & $\sinbb$ \\
      \hline
      $\jpsi\ks(\pip\pim)$    & 1997 & $0.67 \pm 0.08$ \\
      $J/\psi \ks(\piz\piz)$  &  288 & $0.72 \pm 0.20$ \\
      $\psi(2S) \ks$          &  308 & $0.89 \pm 0.20$ \\
      $\chi_{c1}\ks$          &  101 & $1.54 \pm 0.49$ \\
      $\eta_c \ks$            &  217 & $1.32 \pm 0.29$ \\
      \cline{2-3}
      All with $\xi_f = -1$   & 2911 & $0.73 \pm 0.06$ \\
      \hline
      $\jpsi\kl$              & 2332 & $0.80 \pm 0.13$ \\
      $\jpsi\kstarz(\ks\piz)$ &  174 & $0.10 \pm 0.45$ \\
      \hline
      $\ftag = \bz$ ($q=+1$)  & $\nevqp$ & $0.72 \pm 0.09$ \\
      $\ftag = \bzb$ ($q=-1$) & $\nevqm$ & $0.74 \pm 0.08$ \\
      \hline
      $0 < r \le 0.5$         & 2985 & $0.95 \pm 0.26$ \\
      $0.5 < r \le 0.75$      & 1224 & $0.68 \pm 0.11$ \\
      $0.75 < r \le 1$        & 1208 & $0.74 \pm 0.07$\\
      \hline
      data set I (78 fb$^{-1}$) & 3013 & $0.73 \pm 0.07$ \\
      data set II (62 fb$^{-1}$)& 2404 & $0.74 \pm 0.09$ \\
      \hline \hline
      All                     & 5417 & $0.733\pm 0.057$ \\
    \end{tabular}
  \end{ruledtabular}
\end{table}
All values are statistically consistent with each other.
\begin{figure}
  \includegraphics[width=0.6\textwidth,clip]{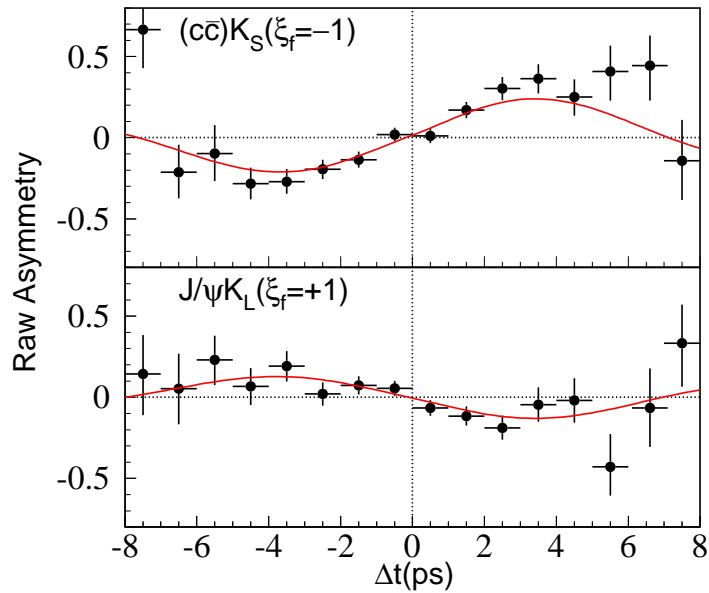}
  \caption{\label{fig:rawasym}
    The raw asymmetries for $(c\overline{c})\ks$ ($\xi_f = -1$) (top) and
    $\jpsi\kl$ ($\xi_f = +1$) (bottom).
    The curves are the results of the global unbinned maximum-likelihood fit.}
\end{figure}
Figure~\ref{fig:rawasym} shows the raw asymmetries
and the fit results for $(c\overline{c})\ks$ (top) and $\jpsi\kl$ (bottom).
A fit to the non-$CP$ eigenstate modes 
$\bz \to D^{*-}\ell^+\nu$ and $\jpsi\kstarz(\kp\pim)$,
where no asymmetry is expected, yields $0.012 \pm 0.013$(stat).

The signal PDF for a neutral $B$ meson decaying into a $CP$ eigenstate
[Eq.~(\ref{eq:deltat})] can be expressed in a more general form as
\begin{eqnarray}
  \label{eq:deltat_general}
  \Psig(\Dt, q, w_l,\dwl) \hspace{10cm} \nonumber \\
  = \frac{ e^{-|\Dt|/\taubz} }{4\taubz}
  \Biggl\{ 1 -q\dwl + q(1-2w_l)
  \Bigl[ \cals \sin(\dM\Dt) + \cala \cos(\dM\Dt) \Bigr] \Biggr\} ,
\end{eqnarray}
where $\cals \equiv 2{\cal I}m(\lambda)/(|\lambda|^2+1)$,
$\cala \equiv (|\lambda|^2-1)/(|\lambda|^2+1)$,
and $\lambda$ is a complex parameter that depends on both
$\bz$-$\bzb$ mixing and on the amplitudes for $\bz$ and $\bzb$ decay
to a $CP$ eigenstate.
The presence of the cosine term ($|\lambda| \neq 1$)
would indicate direct $CP$ violation;
the value for $\sinbb$ reported above is determined
with the assumption $|\lambda| = 1$, as $|\lambda|$ is expected
to be very close to one in the SM.
In order to test this assumption,
we also performed a fit using the expression above with
$a_{CP} \equiv -\xi_f \text{Im}\lambda/|\lambda|$
and $|\lambda|$ as free parameters, keeping everything else the same.
We obtain
\begin{eqnarray}
|\lambda| = \lambdaresult,\nonumber \\
a_{CP} = \sinbbcenter \sinbbstat \textrm{(stat)}, \nonumber
\end{eqnarray}
for all $CP$ modes combined.
This result is consistent with the assumption used in our analysis.

\begin{acknowledgments}
We wish to thank the KEKB accelerator group for the excellent
operation of the KEKB accelerator.
We acknowledge support from the Ministry of Education,
Culture, Sports, Science, and Technology of Japan
and the Japan Society for the Promotion of Science;
the Australian Research Council
and the Australian Department of Education, Science and Training;
the National Science Foundation of China under contract No.~10175071;
the Department of Science and Technology of India;
the BK21 program of the Ministry of Education of Korea
and the CHEP SRC program of the Korea Science and Engineering Foundation;
the Polish State Committee for Scientific Research
under contract No.~2P03B 01324;
the Ministry of Science and Technology of the Russian Federation;
the Ministry of Education, Science and Sport of the Republic of Slovenia;
the National Science Council and the Ministry of Education of Taiwan;
and the U.S.\ Department of Energy.
\end{acknowledgments}



\begin{thebibliography}{99}
\bibitem{footnote:cc}
  Throughout this paper, when a decay mode is quoted,
  the inclusion of the charge conjugate mode is implied.
\end{thebibliography}
\end{document}